**A Model of Synchronization for Self-Organized Crowding Behavior**

Jake J. Xia[1]

**Abstract**
A general model for synchronized crowding behavior is proposed for understanding and forecasting market instability, drawing similarities from examples such as wildebeests' herding and bridge swaying. Interaction among agents is described as a feedback loop linking individual agent's decision and observation, with influence of external forces. Agents adapt to environment and switch between two distinctive states, normal and reactive. When in reactive state, agents become more responsive to observation. Transition from normal to reactive state is typically driven by loss-avoiding motivations. After sufficient number of agents switch to reactive state, crowd synchronization happens. An order parameter is introduced in this paper to quantify the level of synchronization. It is shown that crowd synchronization is a function of percentage of agents in reactive state. Further, such behavior is shown to be driven by the most active agents (with the highest volatility). A tipping point is identified when crowd becomes self-amplifying and unstable. By applying this model, financial bubbles, market momentum and volatility patterns are simulated. This model has potentially broader applications in studying financial markets, public safety, collective animal behavior, social media, and other group dynamics as measurements become possible for agents' states in addition to observations such as prices.

**Significance Statement**

Crowding behavior is an important phenomenon in many fields. In financial markets, it is directly affecting the system stability. Everyone wants to understand how crowding behavior is caused and developed. This work is attempting to propose a general theoretical framework which can unify different empirical explanations.

Drawing from similarities of several examples in biology and physics such as wildebeests' herding and London Millennium Bridge swaying, I proposed a common framework to explain crowding behavior in financial markets. This physical model is different from a typical finance research approach.

The key difference from this model and many previously proposed models is that no probability distribution is assumed, no specific functional dependence of price over time (such as power law) is needed.

[1] Chief Risk Officer and Managing Director of Harvard Management Company, also Research Affiliate at Massachusetts Institute of Technology's Laboratory for Financial Engineering. Address for correspondence: Harvard Management Company, 600 Atlantic Avenue, Boston MA 02210. This manuscript is a revised version of a working paper of 2006.

## I. Introduction

Background

Synchronization is a common phenomenon observed in many different fields. Among a large group of self-organized autonomous agents, each member has similar individual goals but the group as a whole has no shared common goal. These groups are sometimes called leaderless but coherent multi-agent systems in social studies, or self-propelled entities and active matters in physics. In this paper, we use the term "self-organized crowd" as a generic description for any groups with no central organization, and call members in a crowd "agents".

Examples of synchronization of self-organized crowd can be found in animal herding, bridge-swaying with a large crowd on top [Strogatz et al 2005], traffic jams, trends in social media, neural stimulation, and financial market bubbles. Often agents in a self-organized crowd can have spontaneous behaviors such that the whole group is unintentionally synchronized. If interaction between agents exceeds a certain level, the crowd can become unstable. Damages to agents can occur in the process. With external intervention or internal attenuation, the crowd can eventually revert back to its normal state with a low level of synchronization.

In contrast, organized groups have total objectives, clear definition of roles, order-following rules, and usually have leaders. These groups have enhanced power compared with each individual's capabilities, as they typically aim to maximize the total outcome while rendering individual objectives secondary. The goal of this paper, however, is not to debate the pros and cons of organizations. Instead, this paper will focus on studying self-organized crowding behavior.

Agents in a self-organized crowd (e.g. in many capitalist markets) are governed by individual objectives and some basic rules. In normal circumstances, each agent usually does not have complete information, but tries to make rational decisions. Such system is normally efficient and balanced. Each individual tries to optimize for self-interests, which usually leads to a positive outcome for the group. However, the rules governing the crowd may not be well-designed for all circumstances. During panicking situations and if information is not available, agents may revert to instinctive reactions. They are affected by each other's sentiment, and the natural tendency is to believe the majority is right. Animal spirits (such as greed and fear) can overtake agents' rational thinking. Agents' decisions become reflective to their surroundings. When in danger, survival instincts trump all other considerations. This situation forces most agents to adapt to the environment and be short-term focused. As a result, the group outcome will become sub-optimal. In the worst case, a crisis may happen.

In daily life, many systems are designed by assuming a low synchronization of the crowd. For example, banks operate on the basis assuming not all depositors will take money out at the same time. If the crowd is synchronized, bank-run happens and poor liquidity can lead to bankruptcy. Insurance companies operate under similar assumptions. Another example is that our road system is designed by assuming only a small percentage of all cars will drive at the same section of roads at the same time. If the crowd is synchronized, a traffic jam happens. Such diversification or low synchronization is an important assumption in many social systems. Therefore, detecting signs of possible synchronization is critical to avoid system-overload.

The objective of modeling self-organized crowd is to better understand the interaction between agents and vulnerability of the crowd, so that we can monitor the stability of a crowd. This is a difficult problem to solve as behaviors of live agents are hard to predict. One has to deal with uncertainties due to high level of complexity and reflexivity, and lack of information of each individual. At group level, some behavioral patterns are more universal and stable. When certain conditions are met, an eventual outcome can be forecasted, though predicting the exact timing is still not possible.

Review

Crowd behavior was first studied in psychology and sociology. For example, Sigmund Freud in his paper "Group Psychology and Analysis of the Ego" highlighted people's inability to resist emotional contagion in a group. Crowd provides a momentary release of otherwise repressed impulses confined within individuals. Psychologist William McDougall particularly studied self-organized crowd behaviors. He stated that the greater the number of people can be observed with the same emotion, the greater the contagion. Perhaps the most cited work was by Gustave Le Bon in his 1896 book "The Crowd, a study of the popular mind" [Le Bon 1896]. Unconscious action or mania of crowd is due to psychology. For the need of social belonging, people give up their own rational thinking to the crowd thinking. Crowd has a transitory collective mind, as coupling among the crowd converge members to one idea. Losing conscious personality or diversity causes high correlation [Hebb 1949]. On the other hand, the wisdom of the crowds is also used as a better predictor in many statistical measurements, under certain limitations [Lorenz et al 2011].

In biology, collective animal behavior such as bird flocking, insect swarming, fish shoaling or schooling, and animal herding have long been observed and studied. At micro level with bacteria or neurons, similar emergent behavior arises from simple rules that are followed by elements without central coordination. In physics, self-propelled entities or active matters are studied by many researchers [Vicsek et al 1995]. In physical systems, particles are coupled at the micro level. Macro behavior emerge from the micro-level interactions. While in social systems, agents are more likely coupled through the macro properties. Crowd study has also been more active in public safety and traffic planning in recent years. Computer simulations are applied to these complex systems.

Bubbles in financial markets are reoccurring phenomena which attracted many researchers' attention [Campbell et al 1997, Focardi & Fabozzi 2014, Barberis et al 2016], especially after the global financial crisis in 2008. Herd behavior has been studied more in economics and financial markets [Scharfstein & Stein 1990, Shiller 2000, Shleifer 2000, Lo 2005]. Statistical methods are usually used to detect patterns displayed in historical data. Many big data projects apply regression or machine-learning techniques. Price change over time is fit with various distributions. Quantitative models are also used to predict these patterns [Gabaix et al 2003]. For example, Log-Periodic Power Law Singularity (LPPLS) model [Sornette 2002] simulates financial bubbles by studying the time dependence of price, where a hazard rate is linked to a critical time that describes the likely timing of bubble bursting.

In many statistical studies, causal relationships are not easily established. Lack of fundamental understanding can lead to incorrect predictions as false patterns are not distinguished. As lessons learned in weather forecasting, it is important to understand the local interactions [Silver 2012]. Agent-based model attempts to capture the causal relationships in financial markets [Farmer 2002, Bouchaud 2010, Helbing & Balietti 2011]. Schematic diagrams are used to track the exact details of inter-linkage between different players in a specific ecosystem. Outcome under different initial and boundary conditions can be simulated. Such results are practically useful but hard to generalize, as they depend on the specifics of agents' roles and market microstructure.

A generic model is needed to capture the common behavior of self-organized crowds in different applications. Instead of modeling all the specific relationships between the agents, we need to capture the fundamental and universal nature in all self-organized crowds. A theoretical framework is needed to unify various work done in different fields. Cross-disciplinary comparison can deepen our understanding of a specific problem. Once a general framework is established, new directions of possible applications can be explored. With that in mind, I attempted to propose in this paper an intuitive model which builds on top of previous work in various fields, with a specific application for financial markets.

Is it ever possible to predict, and possibly influence, the behavior of the crowds that represent the financial markets? Can we quantify the level of fragility of markets (not timing of bubble bursting which depends on specific triggers)? How does each agent's behavior affect the overall condition? Traditionally, market observations such as time series of prices and covariance are the main measurements used because they are readily available. Going forward, can we collect specific information of market participants to better understand their decision-making process? How are market participants' mindsets affected by their observations? The model presented in this paper is motivated by these questions and intended to bring these ideas in an investment management context.

## II. The Crowd Interaction Model

As discussed earlier, a self-organized crowd has a number of agents who have similar individual goals and constraints to survive and gain (take risk and reward). Agents use historical experience to adapt and extrapolate when making decisions. Their action is determined by observations, external forces, and their own conditions. An agent's tendency of getting ahead of others for survival drives the response function of the agent.

The crowd problem here is defined as N agents interacting with continuous dynamics in an environment where external forces and observations of agents' own actions change over time. From the first principle, the basic dynamic equation is that an agent's action (to buy or sell) is determined by external news which he cannot influence and observations which are influenced by his and others' actions. The result of such interaction is that the agents become indirectly coupled via observations.

Figure 1. Interaction Loop of Agent, Observation and External Force

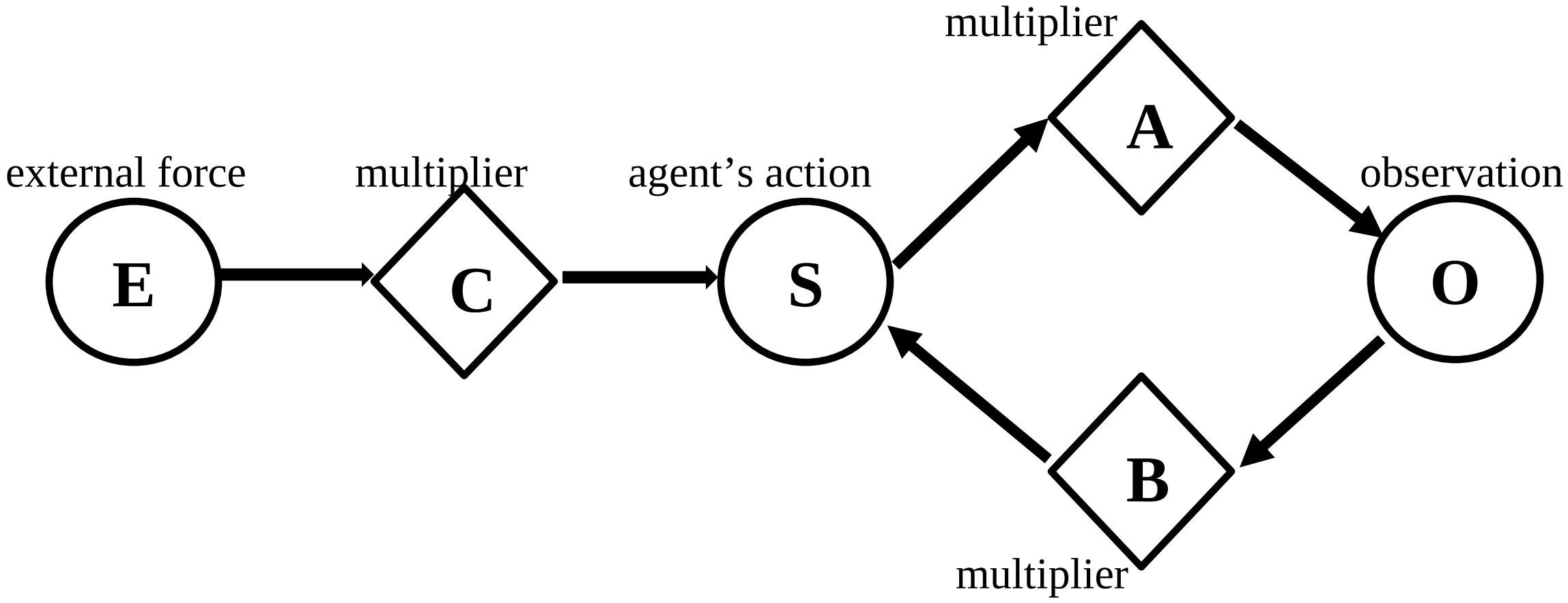

Figure 1 describes the interaction of agents, observation, and external force, which looks similar to the feedback loop in electronic circuits. There are three dynamic quantities in this model: S, O and E. S is the action (to buy or sell) of an agent. Later we will use $S_i$ to denote action of agent i, use S without subscript to denote the aggregated actions of all agents. O is the macro observation that agents see and influence locally. E is the total external force which is not affected by the agent's decision.

These three quantities S, O and E are connected through three response coefficients, A, B and C. A is the sensitivity of observation O to agent's action $S_i$. B is the sensitivity of agent's action $S_i$ to observation O. C is the sensitivity of agent's action $S_i$ to external force E.

In a linear interaction model, A, B, and C are the multiplying constants which determine the amount of impact. In general, these can be any mathematical operators **A**, **B**, and **C** on S, O, and E. They can be non-linear and time-dependent. As we discuss later in different applications, A, B, and C can indeed take different forms operating on S, O, and E. Note that S is determined by E and O, and O is determined by S.

In previous work [Vicsek et al 1995, Helbing et al 2011], interaction was specifically modeled depending on the physical distance. Close neighbor's actions are affecting O and S more than other agents. Here O is generalized to include both local observation influenced by nearby agents and overall observation all agents can see and influence. For example, on a swaying bridge, pedestrians are influenced by the total sway of the bridge as well as the immediate neighbors' behaviors. It is important to point out that the linkage in many self-organized crowds does not depend on physical distance. For example, in financial markets, all participants can observe and influence prices of the whole market, thanks to modern communication technology. Neighbors' actions are actually less important.

dS and dO denote the incremental changes of S and O at a particular time interval t to t+dt. We assume each individual agent i's incremental decision $dS_i$ has different sensitivity or reaction functions $C_i$ and $B_i$ to dE and dO respectively. For simplicity, we further assume A is the same for all agents, i.e. effect on observation is the same for all agents per unit of action. This is a reasonable assumption, though later results do not depend on it. For example, every $ purchase of a stock has the same effect on the price. Additional details can be modeled when special agents' (e.g. leaders') effect needs to be differentiated. Here we simply treat a larger agent as having a bigger number of agent units. N is total number of unit agents in the crowd.

From the flow chart in Figure 1, basic mathematical relationships of the model can be described below.

$$dS_i = C_i \, dE + B_i \, dO + \varepsilon_i \tag{1}$$

In equation (1), $\varepsilon_i$ is a random noise, describing other random decisions by the agents which are not determined by dE or dO. For simplicity, in later discussion, we often assume the impact of $\varepsilon_i$ is small compared with the first two terms and drop $\varepsilon_i$ in modified equation (1'). However, the random noise $\varepsilon_i$ is important when correlation between agents and synchronization are considered.

$$dS_i = C_i \, dE + B_i \, dO \tag{1'}$$

Equation (1') is the basic interacting dynamics that each agent makes an individual decision based on the external force and observation. We also assume that agents' actions are additive,

$$dO = A \sum_{\boldsymbol{i=1}}^{\boldsymbol{N}} dS_i = A \, dS \tag{2}$$

$$dS = \sum_{\boldsymbol{i=1}}^{\boldsymbol{N}} dS_i = \left(\sum_{\boldsymbol{i=1}}^{\boldsymbol{N}} dC_i\right) dE + \left(\sum_{\boldsymbol{i=1}}^{\boldsymbol{N}} dB_i\right) dO = C \, dE + B \, dO \tag{3}$$

$$C = \sum_{\boldsymbol{i=1}}^{\boldsymbol{N}} C_i \tag{4}$$

$$B = \sum_{\boldsymbol{i=1}}^{\boldsymbol{N}} B_i \tag{5}$$

No specific forms of A, B, and C need to be pre-assumed. They are not always constants. Intuitively, agents' actions have the same directional impact on observation, i.e. A>0. Agents usually have a positive aggregated response to external forces, i.e. C>0. Agents' responses to observation can vary, where B can be positive, 0 or negative, depends on agents' aggregated behavior at a particular time. For example, crowd can be inhomogeneous, i.e. $dS_i$ is responding differently depends on the position in the crowd. S and O are functions of location and time. $B_i$ and $C_i$ are different for different agents.

Heterogeneity is usually important in understanding crowd synchronizations. For example, in agent-based models, specific linkages between agents need to be modeled to understand the channels of propagation of market stresses. Some agents may receive additional info of E and O than others. Heterogeneity and randomness also in principle tend to prevent homogeneity and synchronization. In modern financial markets, however, as price observation becomes available to everyone in almost real time, a typical feedback loop between agents' action and observation captures the important macro features of markets. This is similar to mean field theory or effective medium theory in physics such as Curie-Weiss theory of magnetism. Macro properties are captured in these models. The entire system can be simplified to a single feedback loop.

For an organized group, there are usually leaders and organizational rules. A, B, and C are pre-decided (typically from top down) to coordinate and achieve goals for the whole group. Self-organized crowd can adapt to changes by varying B and C over time under different conditions.

Special Case of Instantaneous Response

If the crowd system has zero response time, i.e. the agent can observe its own action instantaneously, we can combine equations (2) and (3) and get

$$dO = \frac{AC}{1-AB}\, dE \tag{6}$$

Under normal conditions, feedback coefficient AB<<1, change of observation, dO, simply responds to the incremental change of external force dE. If AB approaches 1, the crowd reacts violently even when a small change of external force is applied. The crowd will become hyper-sensitive and unstable. AB=1 is a singular point.

General Case of Delayed Response by *dt*

In most practical problems, agents cannot observe their own actions instantaneously, hence the crowd has a delayed response. Equation (2) becomes,

$$dO(t) = A\, dS(t) \tag{7}$$

Equation (3) describes how the agent behaves at *t* + *dt*, based on previously received information at *t*.

$$dS(t+dt) = C\, dE(t) + B\, dO(t) = \; C\, dE(t) + AB\, dS(t) \tag{8}$$

Hence, change of observation at t + dt is related to change of observation at t by

$$dO(t+dt) = AC\, dE(t) + AB\, dO(t) \tag{9}$$

Typically AC>0, observation responds positively to external force. If B>0 and AB>0, the crowd displays a momentum behavior. As long as the external force persists in the same direction or is small comparing with observation, the observed behavior will trend in the same direction. If B=0 and hence AB=0, there is no endogenous reaction, observation simply responds to exogenous signals. If B<0 and AB<0, the crowd has a contrarian behavior counter-balancing to observation.

When there is no external force, dE(t)=0,

$$dO(t+dt) = AB\, dO(t) \tag{10}$$

Here the crowd behavior is simply determined by its endogenous properties, or the feedback parameters A and B. If the absolute value |AB|>1, the crowd can compound on itself and become unstable. In discrete time series, $O(t_n) = (AB)^n dO(t_1)$ , the price displays a power-law

behavior. At each time interval dt, the dO gets amplified by AB times. If AB>1, dO can grow exponentially.

Examples

For a herd of wildebeests, S is the decision to run that determines running speed. O is the average running speed of the group. E is the signal of predators' approaching. Not all wildebeests can sight the lions, hence they also watch others' reaction. When startled, they all run. If B increases value and AB is larger than 1, without lion's attack, the whole herd can self-stimulate and run faster and faster.

For the sway of London Millennium Bridge [Strogatz et al 2005], S is pedestrian's timing of gait that determines the phase difference with bridge sway. O is the magnitude and phase of sway. E is wind or other factors affecting pedestrians. When the sway is strong enough, pedestrians are forced to balance themselves by stepping in sync with the sway, which in turn causes bigger sway. Such sway can be self-induced when AB is greater than 1.

In social media, S is people's decision to join a popular behavior that determines the amount and bias of opinion to post. O is the observed posts. E is the news from outside the social group. When large enough of your "friends" joined a popular movement, you either join them or be left out and become less popular. Your action to join further inflates the popularity.

For financial markets, S is people's decision to buy or sell that determines the amount of transactions. O is the price. E is the news. When $B_i<0$, the player is a contrarian. When $B_i>0$, the player is a momentum follower. When market crashes, investors are forced to cut losses as their financial freedom is threatened. When bubble keeps expanding, fund managers are forced to follow the hypes, otherwise their career is at risk.

There are some similarities between this interactive model and the Hamiltonian function of the Ising model in physics. Aggregated response (or energy) of a lattice of magnetic dipole moments of atomic spins (which can be up or down, +1 or -1) is determined by the coupling of adjacent sites and response to external magnetic fields.

Our crowd model describes the interaction of price and external news. If the random noise is included, Equation (9) can be rewritten in a more general form using equation (1),

$$dO(t+dt) = ACdE(t) + ABdO(t) + A\varepsilon \qquad (11)$$

Where ε is the sum of all $\varepsilon_i$ 's. $A\varepsilon$ is a stochastic process that can be modeled as a Wiener process,

$$A\varepsilon = \mu\, dt + \sigma\, dZ \qquad (12)$$

Where μ is a drift rate parameter, Z is a standard Wiener process or Brownian motion, typically assumed as normal distribution, and σ is the standard deviation or volatility of the time series.

Equation (11) becomes

$$dO(t+dt) = ACdE(t) + ABdO(t) + \mu\, dt + \sigma\, dZ \quad (13)$$

O(t+dt) not only depends on the starting point O(t), but also the history of O(t) before t. Equation (13) combines the stochastic pricing model with the crowd interaction model where the price change depends on the previously change and external news.

Equation (11) describes essentially a first-order autoregressive time series, characterized by a momentum behavior. The crowd model presented here also provides an intuitive explanation of stochastic volatility as described in ARCH/GARCH models [Engle 2001].

## III. Two States of Coupling

With the framework presented in previous section, we next model the coupling coefficients B and C in equation (9), where we will focus on understanding the how B change with O.

Here we introduce an important hypothesis that agents in a crowd switch in between two distinctive states, normal and reactive. In reactive state, coupling coefficients $B_i$ and $C_i$ increase values. Total B and C also increase. In following discussion, we focus on the feedback coefficient $B_i$ of observation. Similar analysis can be applied to the responding coefficient $C_i$ to external forces.

The two-state hypothesis is easy to understand as the underlying drivers are linked with survival instincts. For example, when sensing danger animals switch to reactive state and run for their life. In financial markets, when pain thresholds are reached, investors abandon rational strategies and switch to react. For example, when loss exceeds a significant portion of an investor's wealth, especially when leverage is used, the investor will be forced to sell when market is going down further. When a bubble is expanding, investors will also be forced to buy to stay with peers to protect their jobs. This is another form of stopping out, but on relative basis to peers. Everyone has a limit of loss tolerance. Evolution selected ones with strong survival bias. In normal state, agents pursue different strategies and different time horizons. For example, in financial markets, value investors act like contrarians and hold for longer time than the momentum traders. Balance between various strategies ensures price stability. Higher leverage makes coupling responses stronger. Margin calls or peer pressure can force players to capitulate and focus on short-term survival.

The dual-mode of behavior has deeper psychological roots, as discussed in the work of Kahneman, D. [Kahneman 2013]. People have two distinctive thinking modes: slow and fast. In "normal" state, people take time and apply logical thinking. In "reactive" state, people use experience quickly to extrapolate decisions. In a brain, prefrontal cortex is responsible for slow, complex and logical analysis, while limbic system is for fast, simple and intuitive evaluation, and emotional reactions. Under different conditions, decisions are made by different parts of the brain. The two-state assumption is also very similar to the Ising model in physics.

Figure 2a illustrates the dynamic process when an agent gets close to his pain thresholds. The increased response is shown in a steeper slope of S vs O. B is no longer a constant, it becomes nonlinear and dependent to levels of O and the change dO. Such increased responses exist on both downside and upside of market moves. The reaction decreases after stop-loss levels are cleared. It stabilizes in these hysteresis loops.

Figure 2a. Agent's reaction to observation increases if close to loss threshold

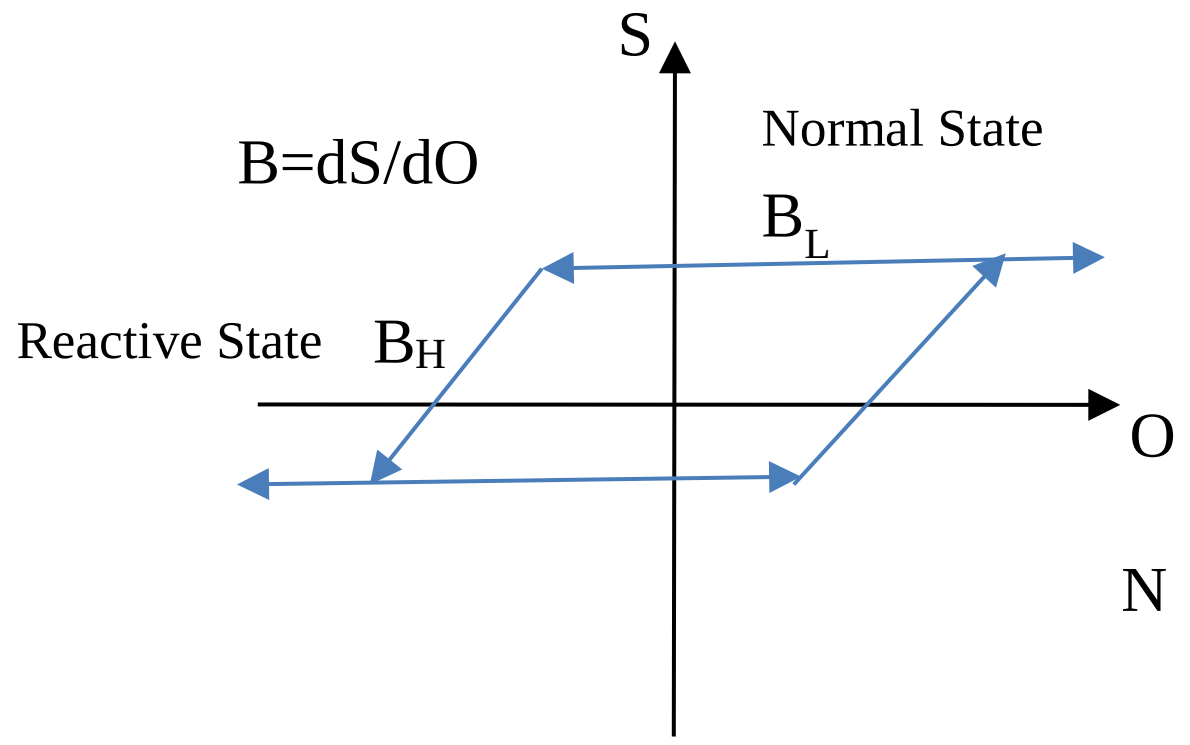

Exact conditions under which an agent switches states depend on the agent's specific situation. Estimation of probability of switch is possible by monitoring the agent's emotional reactions Data can be collected via Twitter, Facebook, social media posting, and questionnaire etc, [Azar & Lo 2016]. In general, the closer an agent is to his pain threshold (in O) and the bigger change of observation (dO), the more likely the agent will switch to reactive state. The focus of this paper is not to study the conditions under which each agent switch states, rather to model the overall impact of such switches.

When in normal state, $B_i$ takes a set of lower values of $\{B_L\}$, indicating low coupling. The set of values $\{B_L\}$ are usually around zero. When in reactive state, $B_i$ takes a higher set of values of $\{B_H\}$, indicating high coupling and agents will try to follow the change in environment when in reactive state. The set of values $\{B_H\}$ are always positive and greater than $\{B_L\}$. Number of agents in reactive state $N_H$, and number of agents in normal state $N_L$ depend on changes of external force E(t) and observation O(t). $N_H+N_L=N$. In simulated results later in this paper, $N_H$ is assumed to be simply proportional to magnitude of dO.

Equation (5) shows that total net feedback parameter B is the sum of all the individual agent's feedback coefficients $B_i$. $B_i$ has two sets of values: low coupling $\{B_L\}$ with $N_L$ number of agents and high coupling $\{B_H\}$ with $N_H$ number of agents. Let $B_H$ be the average of all $\{B_H\}$ and $B_L$ be the average of all $\{B_L\}$. Hence,

$$B = \sum_{i=1}^{N} B_i = N_H\, B_H + N_L\, B_L = N_H\,(B_H - B_L) + NB_L = NB_L[\frac{N_H}{N}(\frac{B_H}{B_L} - 1) + 1] \qquad (14)$$

Equation (14) shows that B is a linear function of $N_H/N$ or percentage of agents in reactive state. B is also proportional to the size of the crowd N and magnitude of $B_H$. $B_H>0$. Typically $B_L<<B_H$ and $B_L \approx 0$. During panic situations, agents switch to high coupling or reactive state. Reactive state may not necessarily mean irrational. It is just the fast reaction. $N_H$ becomes bigger. In turn, the crowd becomes more sensitive to short term observation changes, forming a self-feeding loop. The total crowd coupling coefficient B can vary continuously over time as it depends on $N_H$.

Maximum value of B occurs when $N_H = N$ and $B_{max} = NB_H$. Hence maximum value of $AB_{max} = ANB_H$. Minimum value of B occurs when $N_L=N$ and $B_{min} = NB_L$.

Figure 2b illustrates the two states of $B_i$ and change of total B when $N_H$ changes.

Figure 2b. Two States of Coupling Coefficient $B_i$

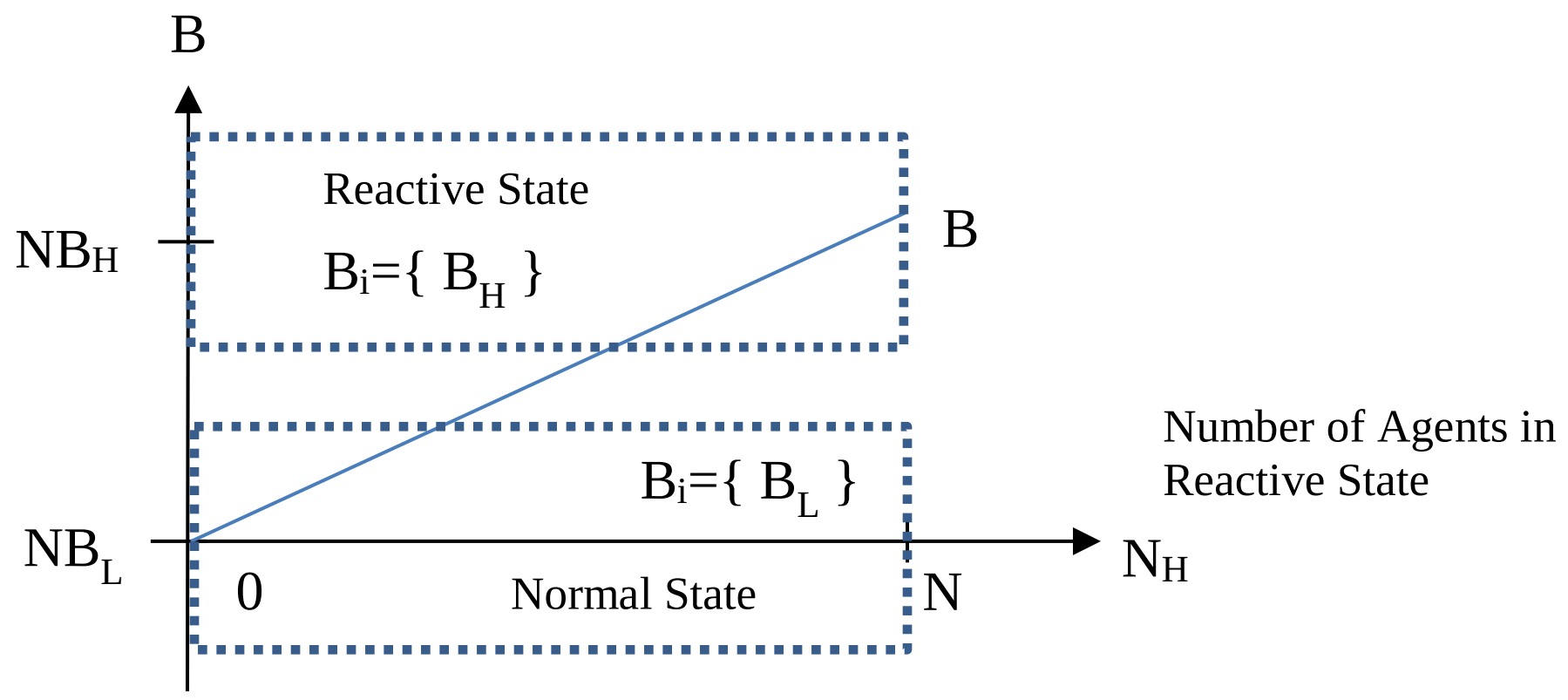

The exact relationship between $N_H$ and dO(t) or dE(t) depends on the specific construct of a crowd. Initial conditions are also important. $N_H$ can be path-dependent on O(t) and may display the hysteresis behavior as in Figure 2a. For each agent, the condition of switching between normal and reactive states is different, as everyone's pain threshold is different. Monitoring agents' state are important in forecasting market behaviors.

## IV. Crowd Synchronization

In this section, the crowd model is applied to study an important phenomenon, synchronization.

At time t, the level of crowd synchronization can be defined as an order parameter,

$$R(t) = \frac{|\sum_{i=1}^{N} dS_i(t)|}{\sum_{i=1}^{N} |dS_i(t)|} \tag{15}$$

0<=R(t)<=1. If all agents move in the same direction, then $dS_i$ have the same sign, R(t)=1. The crowd is synchronized at that time moment t.

From equation (1), when there is no change of external force, dE(t)=0, and no random noise, ε=0. $dS_i(t+dt) = B_i\, dO(t)$.
Using equation (14), we can derive that

$$R(t) = \frac{(\frac{N_H}{N})(B_H - B_L) + B_L}{(\frac{N_H}{N})(B_H - B_0) + B_0} \qquad (16)$$

Where $B_0$ is the average of all absolute values of $\{B_L\}$. Figure 3a shows the relationship between the order parameter R and percentage of agents in reactive state, $\frac{N_H}{N}$.

Figure 3a. Synchronization level increases as more agents switch to reactive state

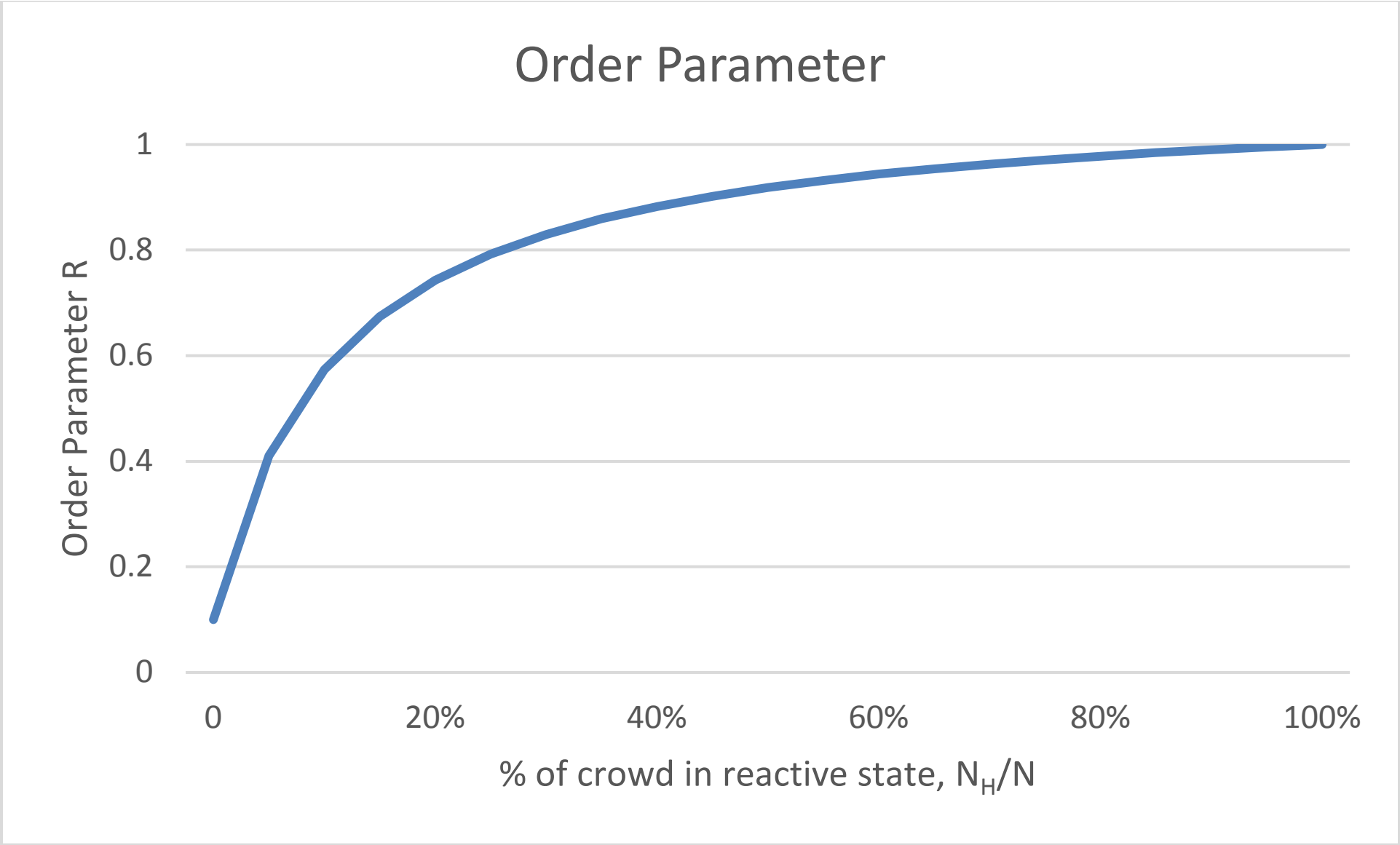

This is essentially the same result as in Vicsek model [Vicsek et al 1995] when noise is set at zero. Similar results are also obtained when we include a noise in equation (1), where R(t) decreases as noise level increases. It is the density of reactive agents (not general agents) that matters to synchronization. Even at zero random noise, the order parameter is usually not at 1 because not all agents are in reactive state. This is an important generalization from Vicsek model.

In Figure 3b, we plotted the impact of random noise on the order parameter for the special case when $N_H = N$. Using Equation (1), $dS_i = C_i\, dE + B_i\, dO + \varepsilon_i$, R(t) in Equation (15) can be computed. The noise $\mathcal{E}_i$ is assumed to be uniformly distributed within values of [-e, +e]. The relative magnitude of noise, e versus $B_H dO$, has an effect on the overall synchronization.

Figure 3b. Synchronization level decreases as random noise increases

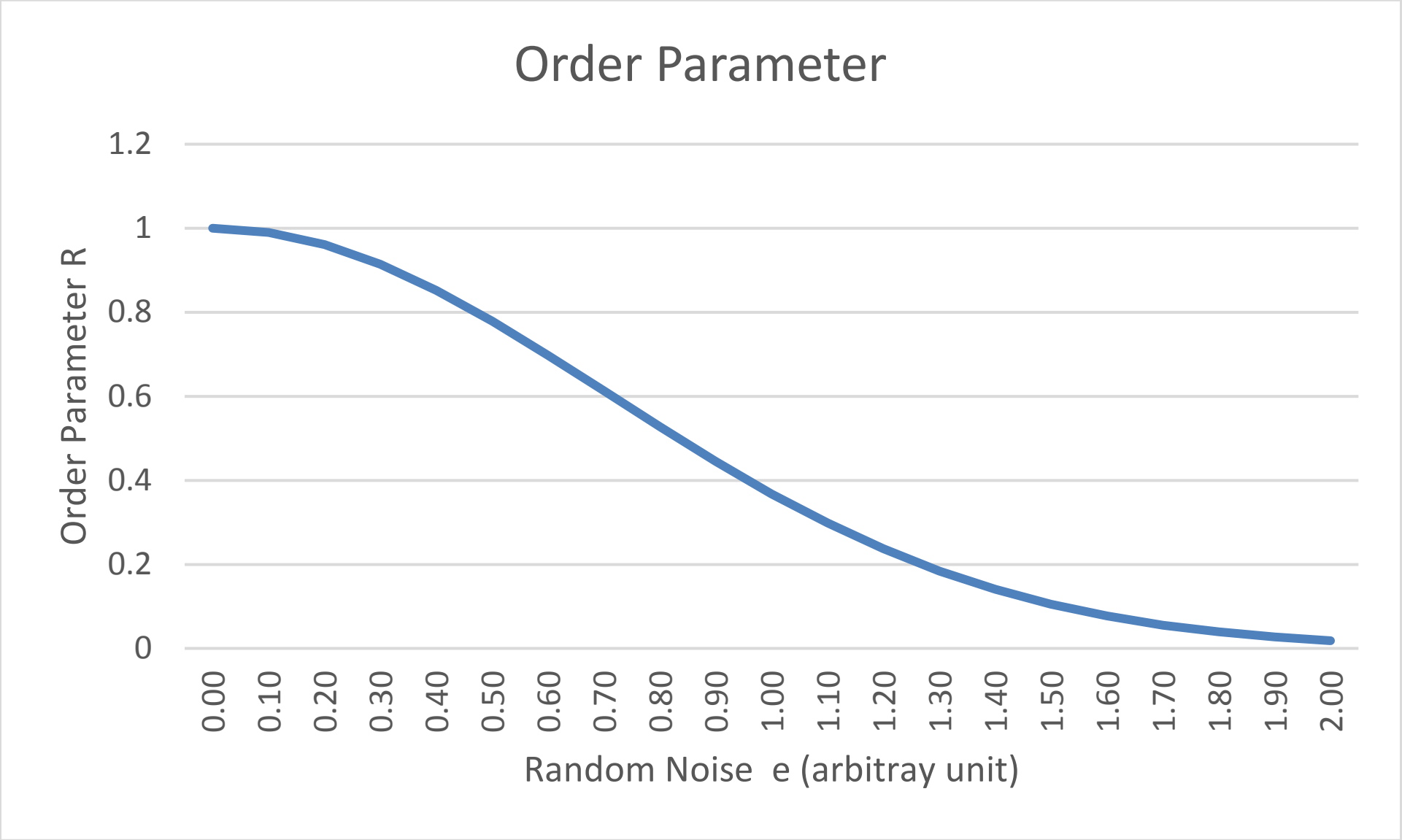

For the sway of London Millennium Bridge [Strogatz et al 2005], O(t) is the horizontal displacement of bridge, S(t) is pedestrians' stepping forces. Strogatz et al. actually solved the mechanical equations to obtain O(t) and S(t), where A, B are nonlinear operators on S and O. Similar order parameter is simulated over time for different crowd sizes, which displays a similar pattern as Figure 3a.

Synchronization level of crowd can also be defined for a time window. As described in equations (1-9), the total crowd's decision dS(t) is the sum of all agents' decisions. Each agent's decision $dS_i(t)$ has a correlation $\rho_i$ to the total decision dS(t) over the time window. The level of crowd synchronization, $\rho_c$, where the subscript c stands for crowd, can be expressed as the average of all agents' correlations to the group's decision,

$$\rho_c = \frac{1}{N}\sum_{i=1}^{N} \rho_i \tag{17}$$

Where

$$\rho_i = \rho[dS_i(t), dS(t)] = \rho\left[dS_i(t), \sum_{j=1}^{N} dS_j(t)\right] \tag{18}$$

Where ρ[X,Y] denotes correlation between two random variables X and Y. It is easy to see that $-1 \leq \rho_c \leq +1$. The higher $\rho_c$, the more synchronized the crowd is.

For any agent i, its decision time series $dS_i(t)$ has a mean, a standard deviation $\sigma_i$, and a correlation $\rho_{ij}$ to another agent j. Correlation matrix $[\rho_{ij}]$is a NxN symmetric matrix. Time series dS(t) is largely influenced by agents with large standard deviations, i.e. large $\sigma_i$'s.

To simply the notations, we will omit time dependence (t) in the equations. Using the definition of correlation, equation (18) can be rewritten as

$$\rho_i = \frac{1}{\sigma_i \sigma_c}\sum_{j=1}^{N} Cov\left[dS_i\,, dS_j\right] = \frac{1}{\sigma_i \sigma_c}\sum_{j=1}^{N} \rho_{ij}\sigma_i\sigma_j = \frac{1}{\sigma_c}\sum_{j=1}^{N} \rho_{ij}\sigma_j \quad (19)$$

Where $\sigma_c$ is the standard deviation of dS, or the volatility of the crowd.

$$\sigma_c{}^2 = Var[dS] = \sum_{l=1}^{N} \sigma_l{}^2 + \sum_{l=2}^{N} \sum_{m=1}^{l-1} 2\rho_{lm}\sigma_l\sigma_m \quad (20)$$

$$\rho_c = \frac{1}{N\sigma_c}\sum_{i=1}^{N}\sum_{j=1}^{N} \rho_{ij}\sigma_j \quad (21)$$

From equation (21), we can see that the overall crowd synchronization depends on correlation matrix between agents. Agents with larger standard deviations (or higher volatility) contribute more to the level of synchronization.

If all agents are perfectly correlated, all $\rho_{ij} = +1$, then, from equation (20) $\sigma_c = \sum_{i=1}^{N} \sigma_i$ .
From equation (21),

$$\rho_c = \frac{1}{N\sigma_c}\sum_{i=1}^{N}\sum_{j=1}^{N} \rho_{ij}\sigma_j = \frac{1}{N\sigma_c}\sum_{i=1}^{N}\sum_{j=1}^{N} \sigma_j = +1.$$

This is the extreme case that all agents in the crowd are perfectly synchronized, typically happens when all agents are in reactive state. In reality, this is unlikely to happen as responses to external forces and observation are different for different agents, and noise terms $\varepsilon_i$ will not be in sync with all agents. Hence, correlation $\rho_{ij} < 1$ and $\rho_c < 1$ .

Equation (21) can be viewed as a weighted sum of the elements of the correlation matrix $[\rho_{ij}]$. Intuitively, the more volatile (or higher amplitude of variation) an agent is, the more important this agent is in measuring synchronization. The correlation of other agents to the volatile agent contributes more to overall synchronization. Note that phase synchronization does not require minimum tracking error, i.e., perfect correlation still allows differences in amplitudes.

In practical applications, since dO(t) reflects dS(t), $\sigma_c$ is therefore observable by measuring standard deviation of dO(t), $\sigma_o$. $\rho_{ij}$ and $\sigma_j$ can only be measured locally with individual agents. Correlation between $dS_i$ and $dS_j$ is defined below where $E[X]$ denotes the mean of random variable X.

$$\rho_{ij} = \frac{Cov[dS_i\,, dS_j]}{\sigma_i\sigma_j} = \left\{E\left[dS_i\, dS_j\right] - E[dS_i]E\left[dS_j\right]\right\}/\sigma_i\sigma_j \quad (22)$$

We can also show that $\rho_c$ increases when $N_H/N$ increases, similar to order parameter R(t). When there is no change of external force and no noise for the whole time window, similar dependence of $\rho_c$ to $N_H/N$ as in equation (16) can be derived.

## V. Bubble Simulation, Tipping Point and Crowd Stability

In this section, the crowd model is applied to simulate bubbles in financial markets under certain external influences. Tipping point of crowd's phase transition is linked with the number of agents in reactive state.

When coupling between agents exceeds a certain level, crowd acts like an amplifier and can become unstable. For example, financial bubbles posts significant systemic risk. Modern technology links in financial markets and people instantaneously. As a result, $B_H$ becomes bigger. A critical point is reached when AB=1. After passing this tipping point, the crowd system becomes self-resonating. External interventions (e.g. by policy makers in financial crisis) are needed to break up the feedback loop.

From equation (14), AB value depends on percentage of agents in reactive state $N_H/N$, size of crowd N, and coupling strength $B_H$ in reactive state. $AB_{max} = ANB_H$. There is a critical point of $N_{HC}$ where AB=1. From equation (14),

$$N_{HC} = \frac{1-ANB_L}{A(B_H-B_L)} = N\,\frac{1-AB_{min}}{AB_{max}-AB_{min}} \tag{23}$$

Since $N_{HC} <= N$, if $AB_{max} <1$, critical point will never be reached.

Financial bubbles are formed when certain market conditions are met. In the crowd model presented in this paper, the key endogenous condition is determined by $AB_{max}$ . External forces act as catalysts. Crowd behavior near critical point is nonlinear. This is understandable since biological systems (including human brain) act very differently when they are near phase transition, as neuron avalanches occur at critical point.

Figure 4. Crowd's response to external force when $AB_{max}$ =0.5

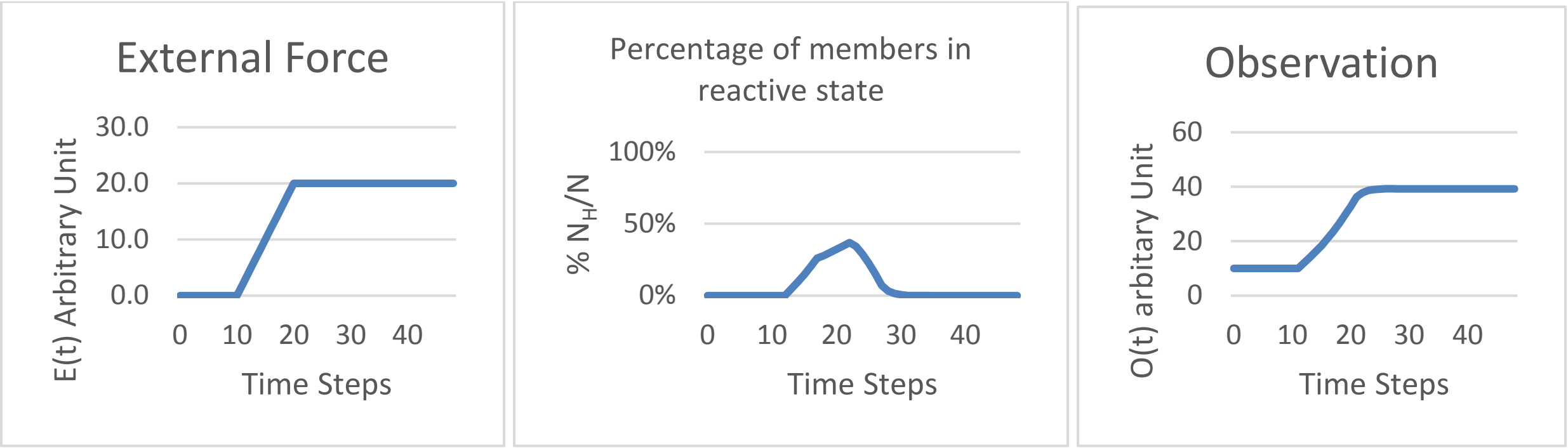

Figure 4 shows an example when $AB_{max}$ <1. An external force is picked to simulate a permanent change of a news, acting as a step-wise function over time. Under the influence of this external force, the crowd responded with 37% of agents switching to reactive state and then all switched back to normal state when external force stabilized. Here we assume $N_H$ is determined by the average magnitude of observation changes (i.e. dO) in the last five time steps.

Figure 5. Crowd's response to external force when $AB_{max}$ =1.35

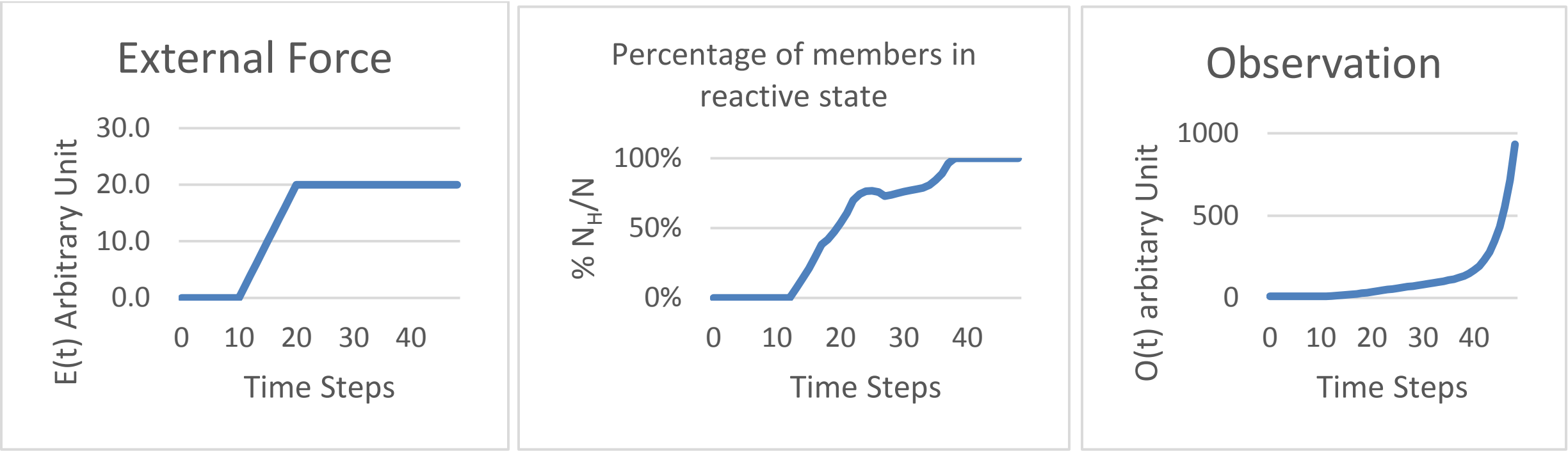

Figure 5 shows an example when $AB_{max}$ >1. Under the same external force as in Figure 4, here 77% of agents fist switched to reactive state. Same as in Figure 4, we assumed that $N_H$ is determined by the average magnitude of observation changes (i.e. dO) in the last five time steps. After external force stabilized, $N_H/N$ only dipped slightly to 73% and then grew to 100% due to self-stimulation which leads to the phase transition. Observation grew exponentially afterwards, crowd became unstable.

Figure 6. Simulated financial bubble growing and crashing when $AB_{max}$ =1

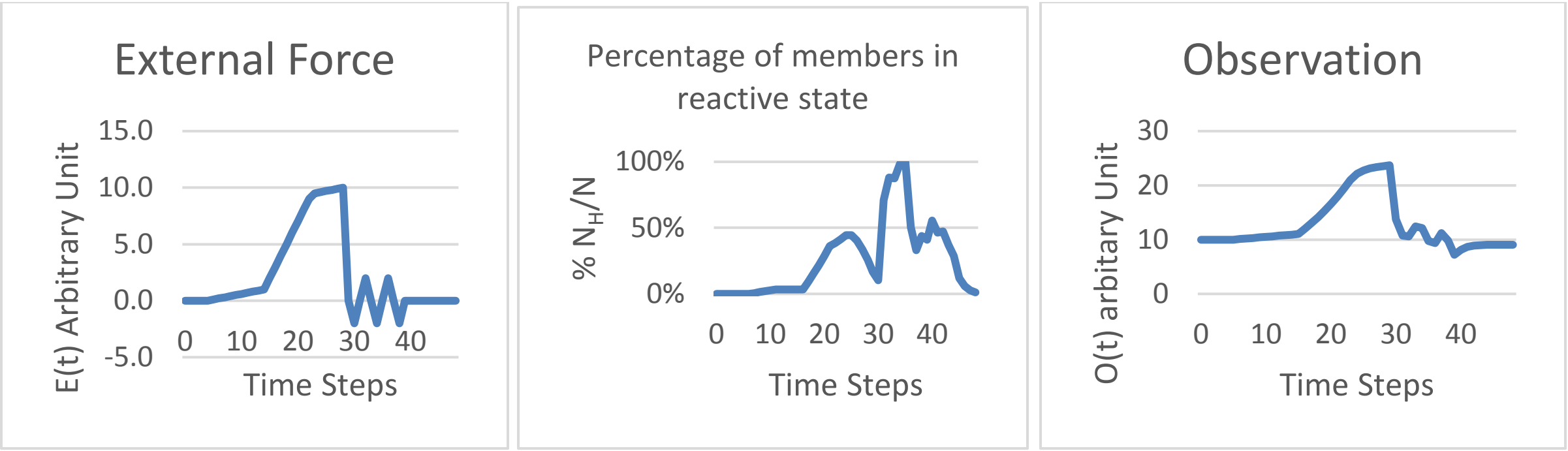

Figure 6 shows the simulated results of a financial bubble induced by external force shown on the left graph. Same as in Figure 4, we assumed that $N_H$ is determined by the average magnitude of observation changes (i.e. dO) in the last five time steps. We picked a typical pattern of an external story which slowly builds up over time, intensifies and positively peaks, then suddenly gets invalidated and follows with confusions. The crowd first responded to the positive change of news with 44% of agents switched to reactive state to chase the bubble. After external good news stalled, $N_H/N$ dropped to 10%. When external news turned negative and became volatile, price (observation) crashed and more agents switched to reactive state with $N_H/N$ eventually peaked 100%. After the external news stabilized, price eventually settled at a lower level. Crowd also became less synchronized again.

## VI. Trend and Volatility Derived from the Crowd Model

This model can also be applied to explain trend and volatility patterns in the market.

The basic idea of a trend is that the direction of change is likely to continue. Trend in a crowd, or the degree of trendiness of a crowd can be quantified by analyzing the time series of agents' total decision S(t) and its reflected observation O(t). Since trend implies monotonic changes in a statistical sense, instead of using numerical methods to find the slope of the best linear fit, we introduce a measurement of trendiness over a time period $t = t_0, t_1, \ldots, t_n$ ,

$$T_d = \frac{|\sum_{l=1}^{n} \Delta O(t_l)|}{\sum_{l=1}^{n} |\Delta O(t_l)|} \tag{24}$$

It is easy to see that $0 <= T_d <= 1$.

When all changes happen in the same direction and all $\Delta O(t_l)$ have the same sign, $T_d=1$, a perfect trend occurs. When all $\Delta O(t_l)$ cancel out and $O(t_n) = O(t_0)$ , $T_d=0$, the crowd has no trend in that time period. When the crowd has a partial trend in that time period, $0 < T_d < 1$.

Trendiness $T_d$ in equation (24) can be computed by simulating dO(t) varying over time in equation (9). As shown in Figures 4-6, trend is stronger if external force continues to change in the same direction and positive feedback $AB_{max}$ is stronger. Since $AB_{max} = ANB_H$, the larger the size of crowd N, and the stronger coupling $B_H$ in reactive state, can lead to stronger possible trend in the market.

Volatility in financial markets is often used as an indicator of panic. Volatility of a crowd, $\sigma_c$ , is observed from volatility of O(t), or $\sigma_O$. From equation (2),

$$\sigma_O = \mathrm{A}\, \sigma_c \tag{25}$$

Equation (20) links $\sigma_c$ with the correlation between agents.

$$\sigma_O{}^2 = A^2 \sum_{l=1}^{N} \sigma_l{}^2 + \mathrm{A}^2 \sum_{l=2}^{N} \sum_{m=1}^{l-1} 2\rho_{lm}\sigma_l\sigma_m \tag{26}$$

Which is a function of individual agent's volatility $\sigma_i$ and correlation matrix $[\rho_{ij}]$. When all agents are in reactive state, total synchronization happens, and all $\rho_{ij} = 1$, $\sigma_c = \sum_{i=1}^{N} \sigma_i$, the crowd volatility $\sigma_c$ reaches the highest possible value given all agents' volatility $\sigma_i$ .

It is practically useful to find a right leading indicator which can forecast market fragility. For example, absorption ratio and turbulence index [Kritzman et al 2011] are derived from variance and correlation of observed price information. In the crowd model discussed in this paper, when agents are in reactive state, volatility and other risk premium move higher. Agents are more risk averse. Loss tolerance is lower. Overall market fragility is linked with the order parameter or $N_H/N$. If we can monitor the pain thresholds (e.g. margin calls levels) of agents and measure the percentage of agents in reactive state, we can have additional information beyond price observations. With possible new methods to collect data, this will open new ways of monitoring financial system stability.

## VII. Discussions

In application of this general crowd model to solve practical problems, the most critical measurements are $N_H/N$ or percentage of agents in reactive state, and $AB_{max}$, the maximum feedback parameter between agents and market observations. This will in turn determine the level of synchronization, crowd order parameter, likelihood of bubble formation and busting, strength of market trend and volatility. In the past, measuring behavior of individual agents locally was very difficult. Thanks to the Internet, social media, and new devices like smart phones, tracking each agent in a crowd becomes possible. For example, apps can be designed to measure how an agent responds to an external signal and to observation changes locally. Many projects demonstrate the power of newly established connectivity. More data will be available to calibrate the model.

Since agents adapt and evolves over time, one should not expect to find a mechanical way to predict the behavior of a crowd. The relationships between $N_H$ (or B) and observation dO(t) and external force dE(t) are not fixed. Agents switch between the states dynamically. Crowd is not a stationary machine. Understanding the specifics of a particular self-organized crowd is absolutely essential. Our general theoretical framework is not intended to replace the specific knowledge of any market. Common behavior across different areas should provide us better insights to understand and predict behavior of various crowds.

Lo's Adaptive Market Hypothesis [Lo 2005] is a good example in finance theory to capture the very nature of ever-changing participants. As Soros stated in his theory of reflexivity [Soros 1987], participants' view of the world is always partial and distorted. These distorted views can in turn influence the situation. For example, fundamentals of a company may be altered by its stock price. The structure of the brain is also a source of distortions. Mistakes and distortions cannot be avoided. It is the goal of this study to better understand the dynamics of crowd interaction and change in psychology. Unlike a pure statistical data processing or a finance model, this model is intended to fill the gap of providing an actual interaction mechanism between agents and their observations.

In future work, specific application in financial markets can be achieved by simulating a particular crowd's behavior in a controlled laboratory environment. For example, some specific conditions of 2008 global financial crisis can be applied. By setting margin call levels and stop-less levels for each agent, one can simulate how agents impact each other through the feedback loops established in this crowd model. By tracking the contagious possibilities, one can run various scenarios analysis and study possible ways to prevent a system crash.

As pointed out in the Introduction, self-organized behavior is studied in many different fields. Further comparing and unifying models are possible. For example, Ising model in physics has been frequently applied to financial markets to capture the local interactions. Possible ways can be found to bridge the crowd model presented in this paper with the Ising model. Autoregressive process described in Equation (11) may be further expanded to establish a concrete link with the ARCH/GARCH models of stochastic volatility.

## VIII. Conclusions

In this paper, a general model for self-organized crowding behavior is proposed. This framework is comprehensive and intuitive. Key features of this model are the following:

1) A general feedback model is introduced to describe the interaction between agents in a crowd. Instead of limiting the interaction to be among the neighbors, all agents are impacted also by observation of overall crowd behavior. This is particularly true in financial markets when technology makes linkage between all agents much greater, and all agents can observe the overall outcome. This model is more general than the Ising model where it emphasizes interactions of physical neighbors. External forces are also included in the interaction model.
2) The agents are assumed to be in two distinctive states, normal or reactive. Agents' coupling coefficients to observation increase when in reactive state. As more agents transition from normal to reactive state, the crowd becomes more coupled.
3) An order parameter is introduced to measure overall crowd synchronization, which we have shown is a function of percentage of agents in reactive state. As the percentage of agents in reactive state increases, crowd synchronization increases. As the random noise of agents' action decreases, crowd synchronization increases. Another way to measure synchronization is also introduced as the average of all agents' correlation to the group's total behavior. Mathematically, we proved that this definition is equivalent to weighted average of elements in the crowd correlation matrix, where the weights are individual volatilities of the agents. Hence, it is proven that agent's contribution to overall synchronization is determined by agent's level of volatility.
4) The model shows that as a crowd is more synchronized, self-induced instability becomes possible. Crowd coupling can act like a self-amplifier. A critical point exists as a threshold of number of agents switch to reactive state. External invention is needed to break up the positive feedback loop and stabilize the crowd.
5) Trends and volatility in financial markets can also be linked with crowd dynamics in this model. A quantitative measure is introduced for trendiness in crowd observation. It is shown that both trendiness and volatility are dependent on percentage of agents in reactive state. By monitoring agents' states in the financial system, we can better forecast volatility and fragility of the markets.

ACKNOWLEDGEMENTS: The author would like to thank the following people (in alphabetic order) for their valuable inputs: Stephen Brown, Georgiadis Dionysios, Doyne Farmer, Campbell Harvey, Mark Kritzman, Martin Leibowitz, Andrew Lo, Jen Lu, Andre Perold, Steve Ross, Andre Shleifer, Didier Sornette, Jeremy Stein, Tomaz Vicsek and Kay Xia.